\documentstyle[pre,aps,twocolumn,amssymb,epsfig,floats]{revtex}
\begin{document}
\draft
\twocolumn[\hsize\textwidth\columnwidth\hsize\csname @twocolumnfalse\endcsname
\title{Differential light scattering:
	probing the sonoluminescence collapse}
\author{G. Vacca,\cite{email} R. D. Morgan, and R. B. Laughlin}
\address{Physics Department and Department of Applied Physics, %
	Stanford University, Stanford, California, U.S.A.}
\date{Copyright 1999 by The American Physical Society, %
	Phys.\ Rev.\ E {\bf 60}, R6303 (December 1999)}
\maketitle
\begin{abstract}
We have developed a light scattering technique based on differential measurement
and polarization (differential light scattering, DLS) 
capable in principle of retrieving timing information 
with picosecond resolution without the need for fast electronics. 
DLS was applied to sonoluminescence, duplicating known results 
(sharp turnaround, self-similar collapse); the resolution was limited by 
intensity noise to about 0.5~ns. 
Preliminary evidence indicates a smooth
turnaround on a \mbox{$\lesssim 0.5$-ns} time scale, and suggests the 
existence of subnanosecond features within a few nanoseconds of the turnaround. 
\end{abstract}
\vskip2pc]

Since Gaitan's seminal work~\cite{gaitan90,gaitan92}, 
significant advances have been made in our understanding 
of single-bubble sonoluminescence (SL). The dissociation hypothesis (DH) 
introduced by Lohse {\em et al.\ }\cite{lohse97} combines the merits of an
intuitive approach based on the relatively tractable Rayleigh-Plesset equation 
(RPE) of bubble dynamics with an impressive ability to reproduce a wide
range of observables~\cite{holt96,matula98,ketterling98,hilgenfeldt99,dan99}. 
More sophisticated theories~\cite{vuong99,moss99} yield a more 
realistic picture of the phenomenon that is in general agreement 
with the results of the DH-RPE treatment.
Experimentally, however, information on the bubble {\em interior}
is still very scarce. In particular, no direct and conclusive evidence exists 
yet for either plasma formation or shock waves inside the collapsing bubble.
As a result, many competing theories~\cite{eberlein96,tornow96,prosperetti97,%
lepoint97,frommhold98,garcia98,mohanty98,willison98} still vie with the
adiabatic or shock-wave heating theory for the distinction of accurately 
describing the SL phenomenon.
It seems desirable, then, to explore additional ways of probing
the interior of the bubble with a time resolution comparable~\cite{szeri99} 
to the duration of the flash, measured to be 
40--380~ps~\cite{gompf97,hiller98,moran98,pecha98}.
Light scattering has already been shown~\cite{lentz95,weninger97,delgadino97} 
to be a useful probe of the bubble dynamics, sensitive as it is to the 
dielectric interface at the bubble wall. It is also a promising
candidate for detection of either plasma or shock waves, since
both features can modulate substantially the local dielectric
constant. 

Our goal was to push measurements of the light scattering cross section
of the collapsing bubble to a greater temporal resolution than that afforded by 
the pulsed Mie scattering technique~\cite{weninger97}, 
which appears to be limited to
around $\frac{1}{2}$~ns by low light levels and the 
need for averaging. 
In order to achieve higher resolution, we have developed a technique 
called differential light scattering (DLS) that reduces statistical 
uncertainty in the detection process by making use 
of more powerful ultrafast laser pulses. 
Since DLS does not rely on a fiducial time reference,
it, too, is completely insensitive to electronic timing noise, 
allowing the use of relatively slow detectors.
The DLS technique is based on two central concepts: (i) using a differential 
measurement to yield jitter-free timing information, and (ii) using 
polarized light to generate such a measurement through
scattering.

\begin{figure}
\begin{center}
\epsfig{file=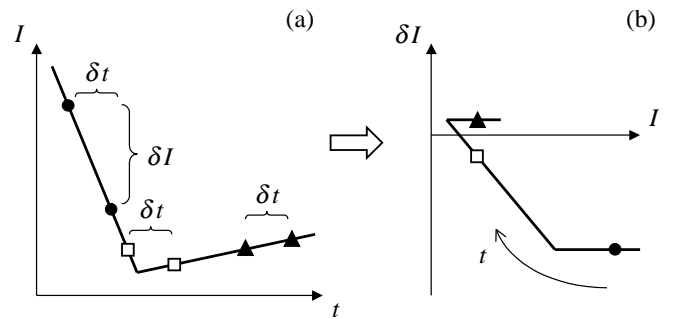,%
	clip=,angle=-90,width=3.375in}
\end{center}
\caption{DLS concept applied to a simple curve. (a) Sampling of $I(t)$ occurs
in pairs; (b) for each pair, the difference signal is plotted against the 
first signal. Equal symbols indicate the correspondence between (a) and (b). 
Notice the slanted ``wall'' in (b) resulting from the two
samples lying on opposite sides of the sharp cusp in (a).} 
\label{fig1:dls}
\end{figure}

The differential measurement concept was recently introduced for the 
first time by Rella~{\em et al.} in the context 
of ultrafast gating of optical 
pulses~\cite{rella98}. The technique they invented, differential
optical gating (DOG), was used to measure the shape of a midinfrared pulse
with subpicosecond resolution. Our technique applies the DOG concept to 
light scattering. DLS relies on collecting many pair
of correlated samples of the same periodic event $I(t)$ 
(see Fig.\ \ref{fig1:dls}).
Each pair $i$ consists of the first sample $I(t_{i})$ 
and the second sample $I(t_{i}+\delta t)$,
where $t_{i}$ is the time of the first sample (modulo the period $T$),
and $\delta t$ is an appropriately chosen
(and short) time delay. From each such pair, an intensity difference 
\begin{equation} \label{eq:Idif}
\delta I_{i}=I(t_{i}+\delta t)-I(t_{i})
\end{equation}
is produced and plotted against the first sample $I(t_{i})$, 
generating what we will call a DLS plot. When enough event pairs
are collected, the points representing them in the DLS plot will join together 
in defining a continuous curve.

The DLS plot can be thought of as a predictor: given an intensity $I$
at some time $t$, the plot yields what $I$ will be after a time
$\delta t$. One can numerically step along the curve on this
plot to retrieve the desired direct function $I(t)$. Depending on the
nature of the features of interest, in some cases it is more fruitful
to plot, for each pair, $\delta I_{i}$ against the second sample
$I(t_{i}+\delta t)$. 
The reconstruction of $I(t)$ is then carried out backwards in time. When the
data are particularly noisy, however, features apparent in the DLS plot
will be lost in the reconstruction process, eliminating any benefit of 
the technique. In this case, it is better to work directly in DLS
space. Using a model for $I(t)$, a DLS curve can be generated 
from it by applying map (\ref{eq:Idif}), then fit to the data points
in the DLS plot with a minimization algorithm.

\begin{figure}
\begin{center}
\epsfig{file=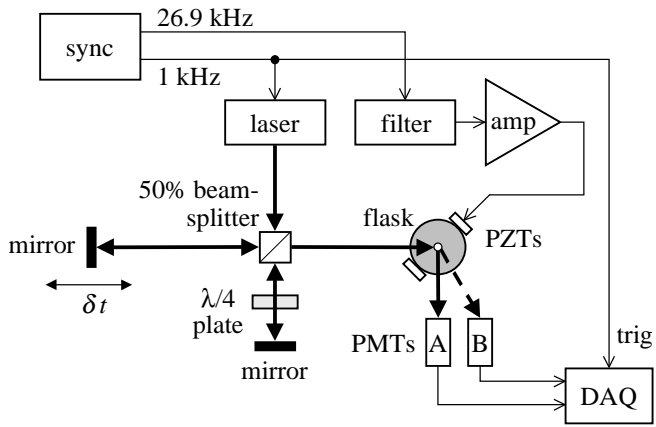,%
	clip=,angle=-90,width=3.375in}
\end{center}
\caption{Experimental setup. Thin lines, electrical
signals; thick lines, beam paths; dashed line, beam going into the page.}
\label{fig2:setup}
\end{figure}

\begin{figure}
\begin{center}
\epsfig{file=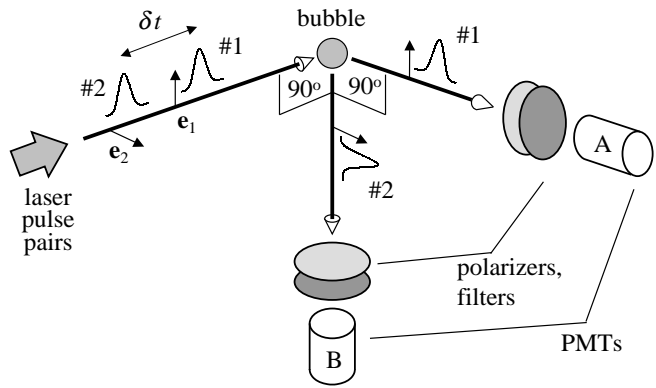,%
	clip=,angle=-90,width=3.375in}
\end{center}
\caption{Scattering and polarization geometry. 
Not shown: lenses focus the input beam onto the bubble and relay
scattered light onto the detectors.}
\label{fig3:polar}
\end{figure}

To implement the DLS concept (Fig.\ \ref{fig2:setup}), 
a laser pulse is split equally in two and recombined as in a 
Michelson interferometer, with one pulse having traveled
a longer path. The two-pulse train is focused onto the bubble, yielding
two bursts of scattered light separated by an adjustable time delay 
$\delta t$. 
In order to distinguish between the two pulses during detection, 
an additional degree of freedom is needed. 
Color discrimination is a possibility, but with several
drawbacks, among which the need for frequency doubling and the strong 
wavelength dependence of light scattering. Polarization, on the
other hand, is perfectly suited to this technique.
Calculations using Mie scattering theory~\cite{hulst81}, which 
is rigorously valid for 
spheres of arbitrary size, show that scattered intensities
are highly polarization dependent.
For the case of linearly polarized light,
scattering at $\theta=90^{o}$ (where zero is forward)
vanishes if the polarization vector {\bf e}
is parallel to the scattering plane. Therefore, the polarization
of one of the pulses is made to rotate by $90^{o}$ (with the 
quarter-wave plate shown in Fig.\ \ref{fig2:setup}) before the two pulses
are recombined.
Figure \ref{fig3:polar} shows the sequence of the two pulses scattering from the bubble. 
The first pulse scatters preferentially in 
the plane containing one photomultiplier tube (PMT), 
while the second pulse does so in the perpendicular
plane, which contains the other PMT. 

Combining the time delay and the polarization dependence results in the
ability to assign the scattered intensity recorded by one detector to the
earlier pulse, and that recorded by the other detector to the later pulse.
Therefore, each laser burst yields an ordered pair of scattered intensities. 
By scanning the arrival time of the pulse pairs over some portion of
the acoustic cycle, enough data
can be collected to generate a DLS plot of the desired interval of
the bubble's evolution.

The experiments were performed in a 100-ml spherical boiling flask
filled with distilled, deionized water 
(resistivity \mbox{$\rho=15.9$~M$\Omega$-cm}) at $20\pm1$~$^{o}$C.
The water was prepared in a gas-handling system under an air pressure
of 0.2~bar, then loaded into
the flask without further exposure to air. A sealed connection
to a volume reservoir (kept at atmospheric pressure for these experiments) 
inspired by \mbox{Ref.\ \cite{hiller95}} provided
pressure release from volume changes induced by temperature fluctuations.
The entire assembly was leak tested; with the flask under vacuum, 
the rate of pressure rise was conservatively determined to be 16~nbar~s$^{-1}$. 
However, under normal operating conditions the flask is filled with water 
and repressurized to 1~bar: this forces outside air to 
diffuse into the undersaturated water through microscopic interfaces, 
resulting in a substantially lower rate of contamination. 
The experiments described here took place 110 days after loading.
Within that time, the pressure in an initially empty flask would have 
risen to roughly 0.2~bar; the air concentration in the water-filled flask 
can instead be expected to have risen by perhaps a few percent 
of atmospheric saturation.

The first acoustic resonance of the flask was determined to be 26.9~kHz.
The acoustic drive was provided by an audio amplifier, the output of which
(typically $\sim 4.5$~W rms)
was fed through an impedance-matching network before being delivered
in parallel to two disc-shaped piezoelectric transducers (PZTs) epoxied 
to diametrically opposite points on the flask. A third, smaller PZT
cemented to the flask provided acoustic pickup, used to map the 
normal modes of the flask and to monitor the behavior of the bubble through its 
filtered acoustic signature.

The laser used to probe the bubble was a regenerative Ti:sapphire 
amplifier pumped by a {\em Q}-switched, frequency-doubled 
neodymium-doped yttrium lithium fluoride (Nd:YLF) laser 
operated at 1~kHz, 10~mJ/pulse (Positive Light 
Spitfire and Merlin, respectively)
and seeded by a 82-MHz mode-locked Ti:Sapphire oscillator, in turn pumped by
an Ar$^{+}$ cw laser (Spectra Physics Tsunami and Beamlok 2080, respectively).
The oscillator provided 60-fs, 800-nm pulses at 82~MHz; the amplifier
output consisted of partially uncompressed (chirped) 50-ps, 800-nm pulses at
1~kHz with approximately 1~mJ/pulse. The dominantly TEM$_{00}$ mode beam 
was sent through a spatial filter to clean up mode asymmetries
and yielded a nearly Gaussian profile. 
To eliminate gross beam distortion caused by the irregular flask surfaces, 
a laser beam input
port was made by cutting a hole in the flask and cementing in place a
custom-made fused silica powerless meniscus.
The light scattered by the bubble was collected with a relay system, 
passed through polarizers (appropriate for each branch) and 800-nm 
narrow bandpass filters, and delivered to two PMTs (Hamamatsu R955P and R636).
The PMT signals were integrated by SRS SR250 boxcar averagers, 
which were in turn sampled by a 1-MHz A/D board on a personal computer.

The synchronization scheme (shown schematically in Fig.\ \ref{fig2:setup}) 
involved generating a logic signal at \mbox{$f_{acous}=26.9$~kHz},
and digitally dividing its frequency by 27 to yield another
logic signal at approximately 1~kHz.
The 26.9~kHz logic signal was filtered
before being fed to the audio amplifier to serve as the acoustic drive,
while the 1~kHz signal was used to trigger the laser
and the data acquisition electronics. 
This ensured that the SL drive signal and the regenerative amplifier
pulse trains would be synchronized to each other to about 1-ns 
precision.
Additional timing circuitry allowed for the delay between the SL flash
(which occurs very nearly at the same point of the acoustic 
cycle, within 0.5~ns of turnaround~\cite{weninger97}) 
and the laser pulse pairs to be varied 
continuously by up to 50~$\mu$s, either manually or automatically.
This allowed us to probe the bubble at any given phase 
of the acoustic cycle.

In order to obtain values for
the ambient bubble radius $R_{0}$ and the acoustic drive amplitude
$P_{a}$, we developed a time-stamp technique
that yielded a time series of scattered intensities $I(t)$
over the whole acoustic cycle. A time-to-amplitude converter (TAC, 
566 EG\&G ORTEC) measured the interval (up to a constant offset) elapsed 
between the arrival of the laser pulse and the SL flash, as signaled by
an additional PMT sensitive to SL light only.
The TAC output was logged through a boxcar along with the 
signal from one of the PMTs used in DLS, and used for time-stamping.
Scattering events were recorded as the delay between the laser
and SL was scanned automatically through a whole acoustic cycle. 

Since the result of this procedure was a time series of intensities,
a calibration was performed to establish a conversion from $I(t)$ to $R(t)$.
This was done using a stroboscopic
imaging system similar to that of \mbox{Ref.\ \cite{tian96}},
except that in our case the drive for the LED was locked 
to the same frequency $f_{acous}$ as that driving the bubble.
We obtained $R_{max}$ by fixing the LED time delay so that the bubble 
was shown on the monitor screen 
at maximum size, and $I_{max}$ from the scattering data. A calculation based 
on Mie theory~\cite{hulst81} 
provided the $I(R)$ map necessary to complete the calibration.

\begin{figure}
\begin{center}
\epsfig{file=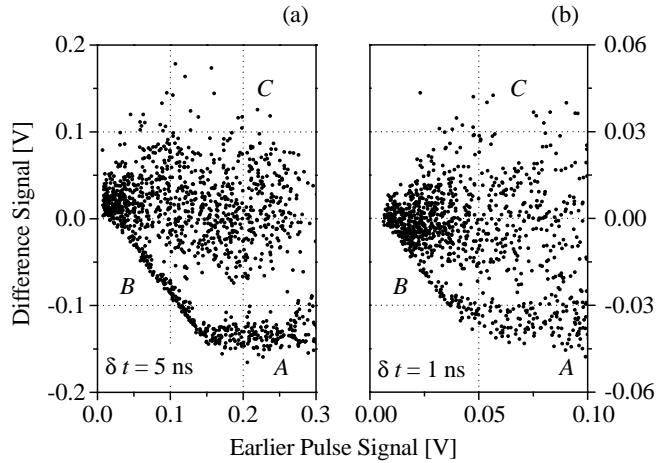,%
	clip=,angle=-90,width=3.375in}
\end{center}
\caption{DLS plots of background-corrected data from a collapsing SL bubble.
Data in (a) is uniformly scaled to account for higher attenuation 
than in (b). Overall range spanned by the
data: (a) 90~ns; (b) 30~ns. The horizontal section {\em A} in (a) shows
self-similar collapse behavior~\protect\cite{delgadino97,barber97,szeri99}
lasting to within 5~ns of turnaround.
The junction between sections {\em A} and {\em B}
appears sharp in (a) but curved in (b), suggesting a smooth turnaround
on a $\lesssim 0.5$~ns time scale. A barely resolved kink in section {\em B} 
of (a) might be due to a subnanosecond feature.}
\label{fig4:data}
\end{figure}

In practice, uncertainties in the calibration of the imaging system,
as well as in the actual measurement of the bubble size, 
prompted us to use our measurement of $R_{max}$ as 
an estimate with $\pm$~10\% uncertainty. The $R(t)$ data were then fed 
to a fitting algorithm that established $R_{0}$, $P_{a}$, and an
appropriate overall scale factor in a nonlinear least-squares 
calculation using the RPE. The fact that the scale factor 
for the best fit was determined
in this way to be \mbox{$1.09 \pm 0.05$} gave us confidence in the 
validity of our imaging method. The bubble parameter values thus found were 
\mbox{$R_{0}=5.3 \pm 0.2~\mu$m} and \mbox{$P_{a}=1.34 \pm 0.04$~bar}.

It is worth mentioning that the time-stamp technique described above
can be used directly to obtain light scattering data from a collapsing
SL bubble. The drawback is that, unlike in DLS, the electronic response of the 
measuring instruments is the limiting factor. We used this procedure to collect
rough timing information as a cross-check in our analysis of DLS data; 
with the devices at our disposal, 2-ns resolution was achieved.
We estimate that with two microchannel plate PMTs, two constant-fraction
discriminators, and a faster TAC, an overall timing uncertainty of 
50~ps should be achievable~\cite{gompf97}.

In Fig.\ \ref{fig4:data} we show representative results 
from our DLS experiments. In these
plots, as in Fig.\ \ref{fig1:dls}~(b), the abscissa is $I(t_{i})$ 
and the ordinate is $\delta I_{i}$.
In Fig.\ \ref{fig4:data}(a) the delay between pulses was 5~ns, 
and in Fig.\ \ref{fig4:data}(b) it was 1~ns.
The range of $\delta t$ for which useful information can be gathered
is dictated by the physical process under study: delays much shorter than
1~ns yielded DLS plots unresolved into a discernible structure, while
delays much longer than 5~ns are not well suited for investigating short
time scales.

To aid interpretation, we divide the plots into three regions:
{\em A} (the collapse, $t<0$), {\em B} (the transition region), 
and {\em C} (the rebound, $t>0$). 
The approximately flat region {\em A} corresponds to the collapse,
since $\delta I_{i}<0$, indicating that the bubble is shrinking. 
In {\em C} the rebounding bubble expands, but at a much lower rate 
than during collapse, so $\delta I_{i}$ is positive and
smaller in magnitude than in {\em A}. However, a greater spread in the data
there results in fuzzy clustering across $\delta I_{i}=0$. 
The straight ``wall'' in region {\em B} forms when the two pulses straddle
$t=0$ (compare to the open squares in Fig.\ \ref{fig1:dls}).

The straight section {\em A} in Fig.\ \ref{fig4:data}(a) is due to
a constant slope of $I(t)$ during collapse. This 
critical behavior~\cite{barber97,szeri99}
has been previously observed for time scales ranging from 1~$\mu$s
to 20~ns prior to turnaround~\cite{delgadino97}; our observations
extend it to $t=-5$~ns. In Fig.\ \ref{fig4:data}(a)
sections {\em A} and {\em B} join rather abruptly, indicating 
a sharp cusp in $I(t)$ on the time scale of the measurement (5~ns).
This was expected given the measurements in Ref.\ \cite{weninger97}.
In Fig.\ \ref{fig4:data}(b), however, section {\em A} appears 
to show a slight upturn before joining section {\em B}, indicating
a smooth transition on a time scale less than the pulse delay of 1~ns
(since the ``wall'' section {\em B} is still discernible).
Scatter in the
data prevents a conclusive interpretation, but the available evidence
would support an estimate of the bubble turnaround time at a few 
hundred picoseconds. 

The collection lenses used have a \mbox{{\em f}-number} of 1.5;
the finite acceptance cone they subtend introduces a
pollution, or cross talk, of unwanted light from the other pulse in each
detector. Because of the strong polarization dependence of 
scattering, this cross talk is quite small: it was calculated from
Mie theory, and confirmed experimentally, to be less than 5\% 
of the total scattered intensity.
Electrical cross talk was measured to be less than 5\%. The resulting
overall intensity uncertainty is therefore around 7\%;
the difference uncertainty varies across the plot.
While in sections {\em A} and {\em B} the error estimates are
consistent with the observed spread in the DLS data, in section {\em C}
the spread is significantly larger.
This has been observed before in Xe-filled bubbles and
ascribed to nonsphericity~\cite{weninger97}; 
such asymmetry is reported here 
as regularly occurring in air-filled bubbles.

In conclusion, we have introduced DLS, a light scattering technique 
based on the DOG~\cite{rella98} concept of differential measurement 
and on sensitivity to polarization that uses intense ultrashort laser pulses 
to bypass the problem of electronic timing jitter. 
The intensity spread in the data is currently the limiting factor 
in the resolution achieved with this technique. 
Effectively, intensity noise is translated into timing noise by the mapping 
that a DLS plot generates.
Accordingly, the resolution in the data shown is approximately 0.5~ns. 
The {\em intrinsic} resolution of DLS, however, is given by the 
laser pulse width used: with our equipment that can be made as low as 500~fs. 
Data collected from a collapsing SL bubble confirm earlier findings 
of a self-similar solution and of subnanosecond turnaround time;
our preliminary results suggest that the turnaround is smooth on a 
time scale of a few hundred picoseconds.

We are very grateful to H. A. Schwettman and the Stanford Picosecond
Free Electron Laser Center for supporting this research.
We also gratefully acknowledge generous equipment loans by J. R. Willison
of Stanford Research Systems. 
We thank B. P. Barber, B. I. Barker, F. L. Degertekin, R. A. Hiller, 
G. M. H. Knippels, G. Per\c cin, S. J. Putterman, C. W. Rella, H. L. St\"ormer, 
and members of the Stanford FEL 
for technical assistance and valuable discussions.

\end{document}